\documentclass[
pra,amsmath,twocolumn,amssymb,notitlepage
]{revtex4-1}

\usepackage{graphicx} 
\usepackage{amsmath} 
\usepackage{xifthen}
\usepackage{physics}

\usepackage{printlen}
\usepackage{todonotes}

\usepackage{hyperref} 
\usepackage[noabbrev]{cleveref} 
\crefname{appendix}{appendix}{appendices}
\Crefname{appendix}{Appendix}{Appendices}

\usepackage{etoolbox} 
\makeatletter
\appto{\appendix}{%
  \@ifstar{\def\theequation@prefix{A.}}%
          {}%
}
\makeatother

\usepackage[load=prefixed,load=abbr,separate-uncertainty=true]{siunitx} 

\usepackage{xcolor}
\definecolor{mmaGreen}{rgb}{0.560181, 0.691569, 0.194885}

\newcommand{\Ham}{\hat{\mathcal{H}}}
\newcommand{\Fin}{\mathcal{F}}
\newcommand{\Rb}{$^{87}$Rb}
\newcommand{\sub}[1]{_{\mathrm{#1}}}
\newcommand{\submath}[2]{_{\mathrm{#1}{#2}}}
\newcommand{\DOne}{$\mathrm{D}_1$}

\newcommand{\DTwo}{$\mathrm{D}_2$}

\newcommand{\e}{\mathrm{e}}
\newcommand{\ii}{\mathrm{i}}
\newcommand{\etal}{\emph{et~al.}}
\newcommand{\ie}{i.e.}

\newcommand{\romnum}[1]{{\romannumeral#1}}
\newcommand{\pcond}[2]{p(#1 \vert #2)}
\newcommand{\redmel}[3]{\mel**{#1}{\abs{#2}}{#3}} 
\newcommand{\para}{{\mkern3mu\vphantom{\perp}\vrule depth 0pt\mkern2mu\vrule depth 0pt\mkern3mu}} 

\makeatletter
\newcommand*\wt[1]{\mathpalette\wthelper{#1}}
\newcommand*\wthelper[2]{%
        \hbox{\dimen@\accentfontxheight#1%
                \accentfontxheight#11.3\dimen@
                $\m@th#1\widetilde{#2}$%
                \accentfontxheight#1\dimen@
        }%
}

\newcommand*\accentfontxheight[1]{%
        \fontdimen5\ifx#1\displaystyle
                \textfont
        \else\ifx#1\textstyle
                \textfont
        \else\ifx#1\scriptstyle
                \scriptfont
        \else
                \scriptscriptfont
        \fi\fi\fi3
}
\makeatother

\newcommand{\ketmix}[1]{\ket*{\smash{\wt{\psi}}(#1)}}

\newcommand{\dash}{^{\prime}}
\renewcommand{\vec}[1]{\vb{#1}}
\newcommand{\hatd}[1]{\hat{#1}^{\dag}}
\newcommand{\creop}{\hatd{a}}
\newcommand{\annop}{\hat{a}}

\newcommand{\wThrj}[6]{
	\mqty(#1 & #2 & #3 \\ #4 & #5 & #6)
	}
\newcommand{\wSixj}[6]{
	\qty{\mqty{#1 & #2 & #3 \\ #4 & #5 & #6}
	}}
	
\newcommand{\pkgfont}[1]{\texttt{#1}}

\begin{document}

\title{Nonlinear Zeeman Effects in the Cavity-Enhanced Emission of Polarised Photons}

\author{Thomas D. Barrett}
\affiliation{University of Oxford, Clarendon Laboratory, Parks Road, Oxford  OX1 3PU, UK}
\author{Dustin Stuart}
\affiliation{University of Oxford, Clarendon Laboratory, Parks Road, Oxford  OX1 3PU, UK}
\author{Oliver Barter}
\affiliation{University of Oxford, Clarendon Laboratory, Parks Road, Oxford  OX1 3PU, UK}
\author{Axel Kuhn}
\affiliation{University of Oxford, Clarendon Laboratory, Parks Road, Oxford  OX1 3PU, UK}


\date{\today}

\begin{abstract}
We theoretically and experimentally investigate nonlinear Zeeman effects within a polarised single-photon source that uses a single \Rb{} atom strongly coupled to a high finesse optical cavity.  The breakdown of the atomic hyperfine structure in the \DTwo{} transition manifold for intermediate strength magnetic fields is shown to result in asymmetric and, ultimately, inhibited operation of the polarised atom-photon interface.  The coherence of the system is considered using Hong-Ou-Mandel interference of the emitted photons.  This informs the next steps to be taken and the modelling of future implementations, based on feasible cavity designs operated in regimes minimising nonlinear Zeeman effects, is presented and shown to provide improved performance.
\end{abstract}

\maketitle

\section{Introduction}

Quantum networks of stationary nodes interlinked by flying photonic qubits are a key goal for quantum information processing \cite{kimble08, fruchtman16}.  A promising candidate for such systems uses single atoms (or ions) strongly coupled to a high-finesse optical cavity to provide an \emph{a priori} deterministic light-matter interface.  Creating atom-photon entanglement is an essential step towards realising remote entanglement and quantum teleportation in such networks, and this is readily achievable by entangling the spin-state of the atom with the polarisation of a photon \cite{[{}][{ Available online at \url{http://mediatum.ub.tum.de/node?id=637078} (last accessed 18/04/18).}]wilk08,wilk07b,ritter12}.  Complete control over this polarised interface requires that these atomic spin states be individually addressable, necessitating an external magnetic field to lift the degeneracy of the magnetic sublevels.  Whilst previous work with these systems has considered these shifted energy levels to be otherwise equivalent to their zero-field counterparts \cite{wilk07c,wilk07b,wilk07,wilk08} here we show that even at the weakest viable field strengths the breakdown of the hyperfine atomic structure can be significant.  We experimentally observe that nonlinear Zeeman (NLZ) effects lead to asymmetric coupling strengths between orthogonally polarised transitions and we provide an investigation and discussion of the implications for polarised single-photon production from these strongly coupled atom-cavity systems.  In the process we resolve a longstanding contradiction between the observed and predicted asymmetries between the efficiencies of polarised photon production \cite{wilk08,wilk07} before presenting how this deeper understanding informs the future direction of the field.

\begin{figure}
\centering
\includegraphics{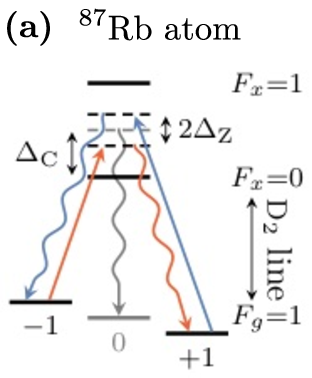}
\includegraphics{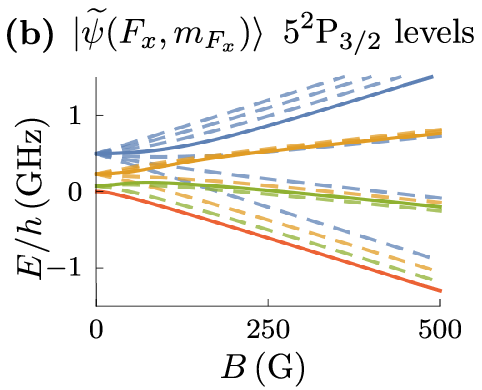}\\
\includegraphics{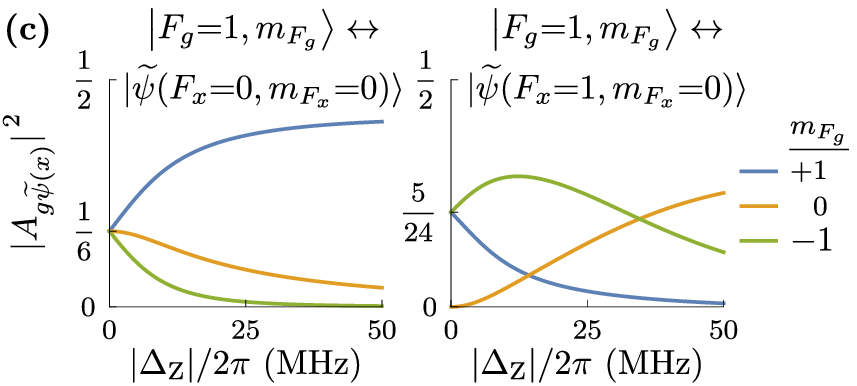}
\caption{(a) Energy level diagram of vacuum-stimulated Raman adiabatic passage (V-STIRAP) between the $\ket{F_g{=}1,m_{F_g}{=}\pm1}$ sublevels of the $5^2\mathrm{S}_{1/2}$ ground level of \Rb{}.  (b) The energy eigenstates of the excited $5^2\mathrm{P}_{3/2}$ level of the \Rb{} \DTwo{} line.  The states that have $\ket{F,m_F{=}0}$ zero-field are shown with solid lines, whilst all other magnetic states are dashed.  (c) Polarisation transition strengths as a function of the ground-state Zeeman splitting, $\Delta\sub{Z}$.  Shown are transitions from the magnetic sublevels of the $\ket{F_g{=}1}$ state to the $\ket{F_x{=}\{0,1\}, m_{F_x}{=}0}$ excited sublevels, expressed as multiples of the \DTwo{} transition dipole matrix element squared, $\abs{\redmel{J_g=1/2}{e\vec{r}}{J_x=3/2}}^2$.}
\label{fig:Theory}
\end{figure}

Several research groups have demonstrated strongly coupled atom-cavity systems as single-photon sources \cite{mckeever04,kuhn02,keller04,vasilev10}, with the high degree of control over the quantum state of the emitted photon extended to its polarisation by Rempe~\etal{} \cite{wilk07c,wilk07} in 2007.  The creation of entangled pairs of photons, emitted sequentially from a single atom \cite{wilk07b}, and of entangled atoms, via a photon emitted by one atom and absorbed by the second \cite{ritter12}, have demonstrated atom-photon entanglement created within such systems.  Our source follows this example, driving vacuum-stimulated Raman transitions on the \Rb{} \DTwo{} line between the stretched states of the $5^2\mathrm{S}_{1/2}$ $\ket{F_g{=}1}$ ground level.  A magnetic field lifts the degeneracy of these levels by $\pm\Delta\sub{Z}$ with respect to the unshifted $\m_{F_g}{=}0$ magnetic sublevel, such that a pump laser alternately detuned by $\pm 2 \Delta\sub{Z}$ from the cavity resonance adiabatically transfers the atomic population from $\m_{F_g}{=}\pm1$ to $\m_{F_g}{=}\mp1$, emitting a $\sigma^{\pm}$ photon into the cavity.  This process is shown along with the atomic structure in \cref{fig:Theory}(a).  The degeneracy of the magnetic sublevels must be sufficiently lifted such that the cavity can selectively couple to either of them.  For a typical cavity linewidth of a few \si{\MHz} this requires $\abs{\Delta\sub{Z}}/2\pi \gtrsim \SI{10}{\MHz}$ (or equivalently field strengths $\gtrsim \SI{14}{G}$).  Whilst this simple consideration holds in the $5^2\mathrm{S}_{1/2}$ ground levels of \Rb{}, the required field strength leads to a breakdown of the atomic hyperfine structure in the $5^2\mathrm{P}_{3/2}$ excited levels.

The breakdown of the atomic hyperfine structure in alkali metals is discussed by Trembley \etal{} in \cite{tremblay90} and the hyperfine Paschen-Back regime of $^{85}$Rb and \Rb{} has since been studied extensively  \cite{sarkisyan04,weller12,hakhumyan12,sargsyan14,sargsyan15}.  The behaviour of three-level systems, in particular when demonstrating electromagnetically induced transparency (EIT), in these strong magnetic fields is presented in \cite{sargsyan12,sargsyan14b,whiting15,whiting16}.  To the best of the authors knowledge the only previous consideration of analogous effects in coupled atom-cavity systems or single-photon sources is that of light shifts caused by an optical dipole trap running orthogonal to the cavity axis found in \cite{neuzner15}.

Here we present the unison of these NLZ effects with a polarised atom-photon interface and show that the complete picture is essential to understand the operation of such systems and to achieve a reliable design in the future.
\section{Theory}

\subsection{Atoms in magnetic fields}

We consider the hyperfine structure of \Rb{} in the presence of a static magnetic field.  For an atom with total orbital angular momentum $\vec{J}$ and nuclear angular momentum $\vec{I}$, the hyperfine Hamiltonian is \cite{steck87}
\begin{equation}
	\Ham\sub{hfs} = A\sub{hfs}\vec{I}\cdot\vec{J} + B\sub{hfs} \frac{ 3(\vec{I}\cdot\vec{J})^{2} + \frac{3}{2}\vec{I}\cdot\vec{J} - I(I+1)J(J+1) }{ 2I(2I-1)J(2J-1)}
	\label{eq:hamHfs}
\end{equation}
where $A\sub{hfs}$ and $B\sub{hfs}$ are the magnetic dipole and electric quadrupole constants respectively.  The interaction of this atom with a magnetic field, $\vec{B}$, is described by the magnetic Hamiltonian
\begin{equation}
	\Ham\sub{B} = \frac{\mu\sub{B}}{\hbar} \qty(g\sub{S}\vec{S} + g\sub{L}\vec{L} + g\sub{I}\vec{I})\cdot\vec{B}.
	\label{eq:hamB}
\end{equation}
where $\mu\sub{B}$ is the Bohr magneton and $\vec{S}$ and $\vec{L}$ are the spin and orbital angular momenta that sum to $\vec{J}$.  For sufficiently weak magnetic fields such that $\expval*{\Ham\sub{hfs}} \gg \expval*{\Ham\sub{B}}$ (or equivalently $A\sub{hfs} \gg \mu\sub{B}\abs{\vec{B}}$) the energies of magnetic sublevels changes linearly with $\abs{\vec{B}}$ and $m_F$, the projection of the total atomic angular momentum $\vec{F}=\vec{I}+\vec{J}$ onto the field.  In the Paschen-Back regime, with $\expval*{\Ham\sub{hfs}} \ll \expval*{\Ham\sub{B}}$ ($A\sub{hfs} \ll \mu\sub{B}\abs{\vec{B}}$), the interaction of $\vec{I}$ and $\vec{J}$ with the external field decouples them from each other and the atomic states are then described by the quantum numbers $\{I,J,m_I,m_J\}$.  These two extreme cases represent standard textbook knowledge.  Here, for the excited $5^2\mathrm{P}_{3/2}$ level of the \Rb{} \DTwo{} line which we primarily consider, the magnetic dipole constant is $A_{\mathrm{hfs,}5^2 \mathrm{P}_{3/2}}/\hbar = 2\pi{\times}\SI{84.72}{\MHz}$ whilst the hyperfine splitting of the ground $5^2\mathrm{S}_{1/2}$ level is much larger at $A_{\mathrm{hfs,}5^2\mathrm{S}_{1/2}}/\hbar = 2\pi{\times}\SI{3.42}{\GHz}$ \cite{steck87}.  For the typical field strengths required in our system the $5^2\mathrm{S}_{1/2}$ level is in the linear Zeeman regime, however the $5^2\mathrm{P}_{3/2}$ level is between the two extremes and so is subject to state-mixing that gives rise to numerous associated non-linearities.

For intermediate strength fields one must generally diagonalise $\Ham\sub{hfs}+\Ham\sub{B}$ and in doing so define the new basis of eigenstates \cite{tremblay90}
\begin{equation}
	\ketmix{F,m_F} = \sum_{F\dash} c_{FF\dash}(\abs{\vec{B}})\ket{F\dash,m_{F\dash}}
	\label{eq:mixedEigenstates}
\end{equation}
where $c_{FF\dash}(\abs{\vec{B}})$ are mixing coefficients.  The sum and mixing coefficients are functions of only $F$ as conservation of angular momentum dictates that only states of the same $m_F$ are coupled by the magnetic field, hence $m_F=m_{F\dash}$.  The nonlinear transition between the weak- and strong-field regimes can be seen in \cref{fig:Theory}(b) which shows a Breit-Rabi diagram of the changing energy eigenstates of the excited $5^2\mathrm{P}_{3/2}$ level as a function of the applied field strength.

When the mixing of hyperfine levels is negligible (\ie{} $c_{FF\dash}(\abs{\vec{B}}) \approx \delta_{FF\dash}$) the transition dipole moment between two magnetic sublevels, $\ket{g} \rightarrow \ket{x}$, can be expressed as
\begin{equation}
	\mel{g}{e r_q}{x} = A_{gx} d,
	\label{eq:dipoleMoment}
\end{equation}
where $d$ is the transition dipole matrix element which, for the \DTwo{} line, is $d=\mel{J_{g}{=}1/2}{\abs{e\vec{r}}}{J_{x}{=}3/2}=\SI{3.58e-29}{\coulomb\meter}$ \cite{steck87}, and $A_{gx}$ is a pre-factor accounting for the angular dependence of the transition coupling strength being considered.  Factoring out the angular dependence in this way is an application of the Wigner-Eckart theorem \cite{eckart30}.  The values of $A_{gx}$ in the absence of a magnetic field can be found in many standard reference books \cite{steck87} and is given by \cite{tremblay90,sargsyan17}
\begin{equation}
	\begin{aligned}
	A_{gx}\left(F_g,F_x,\dots\right) &= (-1)^{1+I+J_x+F_x+F_g-m_{F_x}}\\
	& \times \sqrt{(2F_g+1)(2F_x+1)(2J_g+1)}\\
	& \times \wSixj{J_g}{J_x}{1}{F_x}{F_g}{I_g} \wThrj{F_x}{1}{F_g}{m_{F_x}}{q}{-m_{F_g}}.
	\label{eq:AcoeffZeroField}
	\end{aligned}
\end{equation}
where the terms in braces are the Wigner 3-j and 6-j symbols and $q$ indexes the spherical component of $\vec{r}$ such that the 3-j symbol, $\qty(\,\dotsm)$, vanishes unless ${m_{F_g} = m_{F_x} + q}$.  These coupling coefficients in a magnetic field are then~\cite{tremblay90,sargsyan17}
\begin{equation}
	A_{\wt{\psi}(g)\wt{\psi}(x)} = \sum_{F_g\dash,F_x\dash} c_{F_x F_x\dash}(\abs{\vec{B}}) A_{gx}\left(F_g\dash,F_x\dash,\dots\right) c_{F_g F_g\dash}(\abs{\vec{B}}).
	\label{eq:couplingCoeffsFull}
\end{equation}

Due to the large spacing of the hyperfine structure of the ground $5^2\mathrm{S}_{1/2}$ level of the \Rb{} \DTwo{} line, we can neglect the mixing of ground-state levels for the field strengths we consider, and so have couplings of the form
\begin{equation}
	A_{g\wt{\psi}(x)} = \sum_{F_x\dash} c_{F_x F_x\dash}(\abs{\vec{B}}) A_{gx}\left(F_g,F_x\dash,\dots\right).
	\label{eq:couplingCoeffs}
\end{equation}
\Cref{fig:Theory}(c) illustrates the change in coupling strengths (which are proportional to $\abs*{A_{g\wt{\psi}(x)}}^{2}$) of the ground magnetic sublevels $\ket{F_g{=}1,m_{F_g}}$ to the excited sublevels we primarily couple to in our single-photon source.  As an external field is required to lift the degeneracy of the ground sublevels we express our field strength in terms of the Zeeman shift, $\Delta\sub{Z}$, it provides,
\begin{equation}
	\hbar\Delta\sub{Z} = m_F g_F \mu\sub{B} \abs{\vec{B}}.
	\label{eq:zeemanSplittingWeakField}	
\end{equation}
For $\abs{\Delta\sub{Z}}/2\pi\approx\SI{10}{\MHz}$, it is evident that the transition strengths deviate substantially from the those of the unperturbed atom.

\subsection{Atom-Cavity coupling}

We consider the atom-cavity system to have the basis of eigenstates $\ket{i,n_+,n_-}$ where $i$ is the atomic state and $n_{\pm}$ is the photon number in the $\sigma^{\pm}$ mode of the cavity.  We are operating in the singe-photon regime and so consider the photon number in each mode to be zero or one.  The Hamiltonian of our system can then be expressed as the sum of that for the bare atom, $\Ham\sub{atom}$, the cavity, $\Ham\sub{cav}$, and interaction provided by the laser and cavity couplings, $\Ham\sub{int}$,
\begin{equation}
	\Ham{} = \Ham{}\sub{atom} + \Ham{}\sub{cav} + \Ham{}\sub{int}.
	\label{eq:ham}
\end{equation}
Using a dressed-state basis in the rotating frame with the atom in the lowest energy level of the excited manifold (\ie{} the level with the lowest value of $F_x$) and no photons in the cavity ($n_+=n_-=0$) defined as origin of the energy scale, the bare Hamiltonian is given by
\begin{equation}
	\Ham\sub{atom} = \hbar{} \sum_i \Delta_i \ketbra{i},
	\label{eq:hamAtom}
\end{equation}
where $\hbar\Delta_i$ is the shift of the dressed atomic state $\ket{i}$ from the chosen zero-energy state.  The cavity,  detuned by $\Delta\sub{C}$ from resonance with the transition between $\ket{F_g{=}1,0,0}$ and the zero-energy state, provides the static Hamiltonian
\begin{equation}
	\Ham\sub{atom} = \hbar\Delta\sub{C} \qty( \creop_+\annop_+ + \creop_-\annop_- ),
	\label{eq:hamCav}
\end{equation}
where $\creop_{\pm}$ is the creation operator for a photon in the $\sigma^{\pm}$ mode.

For geometric reasons the cavity, which is aligned along the same axis as the magnetic field, cannot support $\pi$-polarised light and only couples transitions with $\Delta m_F = \pm1$.  Transitions coupled by the cavity require either the emission or absorption of single photon and so change both the photon number and the atomic state.  Atomic transitions with $\Delta m_F = \pm1$ from the ground to excited state are associated with the absorption ($\annop_{\pm}$) and emission ($\creop_{\pm}$) of a $\sigma^{\pm}$ photon by simple conservation of angular momentum argument.  The laser is linearly polarised perpendicular to the axis of the magnetic field and injected orthogonally to the cavity such that it decomposes into a superposition of $\sigma^{\pm}$ light in the atomic basis.  As such it also couples only $\Delta m_F = \pm1$ transitions within the atom whilst leaving the photon number in the cavity unchanged.  The interaction Hamiltonian is then
\begin{equation}
	\begin{split}
    \Ham\sub{int} = - \hbar \sum_{i,j} & \Big\{ \tfrac{1}{2}A_{ij}\overline{\Omega}\qty(t) ( \ketbra{j}{i} + \ketbra{i}{j} )\\
      & + A_{ij}\overline{g} ( \ketbra{j}{i}\annop_{\pm} + \creop_{\pm}\ketbra{i}{j} ) \Big\},
	\end{split}
	\label{eq:hamInt}
\end{equation}
where $i$ and $j$ index the coupled atomic states and $\creop_{\pm}$, $\annop_{\pm}$ are creation and annihilation operators for photons of the appropriate polarisation.  Once again we factor out the angular dependence of the transition dipole moments as seen in \cref{eq:dipoleMoment} with the atom-cavity coupling rate and the Rabi frequency of the driving laser respectively given by
\begin{equation}
	g=A_{ij}\overline{g}, \qquad \Omega\qty(t)=A_{ij}\overline{\Omega}\qty(t).	
\end{equation}
This definition of the barred coupling rates, $\overline{g}$ and $\overline{\Omega}\qty(t)$, leaves them dependent only on the transition dipole moment of the atom and physical parameters of the light-field.  Importantly they are independent of the specific magnetic sublevels. 

The time evolution of this system is described by the master equation \cite{brasil13,pearle12},
\begin{equation}
	\frac{\diffd}{\diffd t} \hat{\rho} = -\frac{\ii}{\hbar}\comm{\Ham}{\hat{\rho}} + \hat{L}(\hat{\rho}),
	\label{eq:masterEquation}
\end{equation}
where $\hat{\rho}$ is the density matrix and $\hat{L}(\hat{\rho})$ is the Liouville operator accounting for the relaxation of the system.  It takes the form
\begin{equation}
	\hat{L}(\hat{\rho}) = \sum\limits_{n} \left( 2\hat{C}_n \hat{\rho} \hatd{C}_n - \hat{\rho}\hatd{C}_n\hat{C}_n - \hatd{C}_n\hat{C}_n\hat{\rho} \right)
	\label{eq:liouvilleOperator}
\end{equation}
with $\hat{C}_n$ the collapse operators.  In our case these operators are either photon emission from the cavity,  with $\sqrt{2\kappa} \annop_{\pm}$, or spontaneous decay between atomic levels $i \rightarrow j$ with $\sqrt{2\gamma_{ij}} * \ket{s_{j}}\bra{s_{i}}$, where $\kappa$ and $\gamma_{ij}$ are, respectively, the cavity field decay rate and atomic amplitude decay rate between levels $i$ and $j$.

\subsection{Simulations}

The simulations presented in this work are performed in two steps.

\begin{enumerate}
	\item Diagonalisation of $\Ham\sub{hfs}+\Ham\sub{B}$ is performed using the \pkgfont{AtomicDensityMatrix} package in Mathematica \footnote{Developed by Simon Rochester and available for downloaded from \url{http://rochesterscientific.com/ADM/} (last accessed 11/01/17).}.  The full excited manifold of the \DTwo{} (\DOne{}) line is used to calculate the energy shifts and coupling strengths in the chosen magnetic field.
	\item The shifted energy levels and modified coupling strengths are put into the Hamiltonian for the system, \cref{eq:ham,eq:hamAtom,eq:hamCav,eq:hamInt}, with all the magnetic sublevels of the $\ket{F_g{=}1}$ ground level and the $\ket{F_x{=}\{0,1\}}$ ($\ket{F_x{=}\{1,2\}}$) excited levels of the \DTwo{} (\DOne{}) line.  The master equation, \cref{eq:masterEquation}, is then solved numerically using the \pkgfont{Qutip.mesolve} \cite{johansson13} Python \pkgfont{package.function} to simulate the process.
\end{enumerate}
\section{Experimental Setup}

\begin{figure}[b]
\centering
\includegraphics{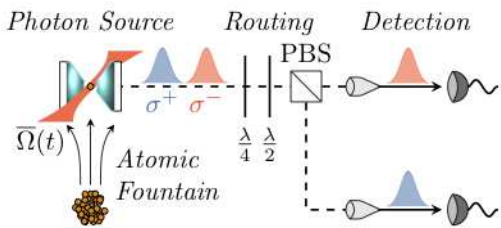}
\caption{Experimental set-up of the atom-cavity source, showing the production, routing and detection of polarised single photons.}
\label{fig:ExperimentalSetup}
\end{figure}

The experimental set-up is shown in \cref{fig:ExperimentalSetup}.  \Rb{} atoms are loaded into a magneto-optical trap (MOT) ${\sim}\SI{8}{\mm}$ below the cavity for $\SI{500}{\ms}$.  They are then stochastically loaded by an atomic fountain -- an upwardly launched MOT with sufficiently diffuse density upon reaching the cavity that in general only one or zero atoms are loaded at any given time.  As the atomic cloud transits the cavity a sequence of \num{25000} driving pulses, alternating in detuning by $\pm 2 \Delta\sub{Z}$, attempts to produce a stream of alternately polarised single-photons.  Opposite polarisations of $\sigma^{\pm}$ photons are directed to different  paths by standard polarisation routing optics.  Superconducting nanowire detectors~\footnote{Photon Spot, model number NW1FC780.} detected these photons with efficiencies exceeding \SI{80}{\percent} with every event recorded at run time with by a commercial time-to-digital converter~\footnote{Qutools quTAU.}.

The cavity itself is \SI{339}{\um} long and is comprised of two highly reflective mirrors with $R\sub{cav}=\SI{5}{\cm}$ radii of curvature and differing transmissions at \SI{780}{\nm} of approximately \SIlist{4;40}{ppm} for directional emission of the photons.  The finesse of the cavity is $\Fin{}{=}\num{118000}$ with a linewidth of $\Delta\omega\sub{FWHM}/2\pi=\SI{3.75}{\MHz}$.  The parameters used when simulating this system are then $\{\overline{g}{=}0.7\,\overline{g_0},\kappa,\gamma\}/2\pi = \{7.32,1.875,3\}\,\si{\MHz}$, where the atom-cavity coupling, calculated using the transition dipole matrix element of the \Rb{} \DTwo{} line, is set to \SI{70}{\percent} of the theoretical maximum value, $\overline{g_0}$, to account for the variations in coupling experienced by atoms in free flight through the standing-wave profile of the cavity mode.  Correcting for this effect by considering a reduced coupling follows the example of previous work within the field \cite{vasilev10,nisbet13,dilley12b,wilk07}.
\section{Results}

\subsection{Observing NLZ effects}

\begin{figure}
\includegraphics{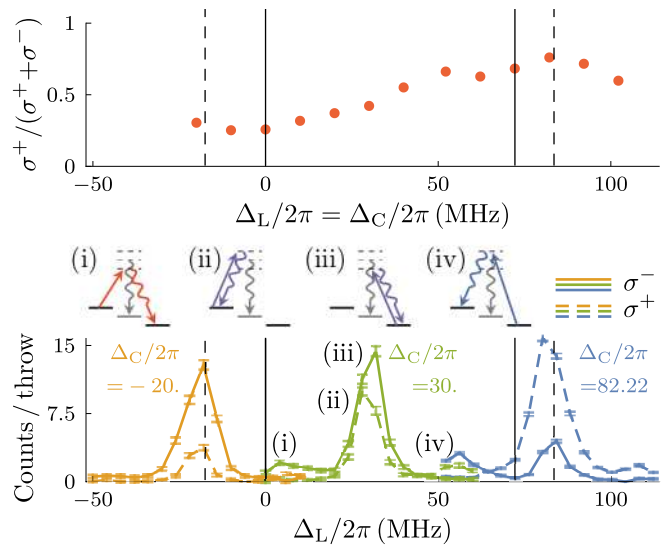}
\caption{The top plot shows the relative emission rates of $\sigma^{+}$ and $\sigma^{-}$ photons of a resonantly driven atom-cavity system.  The Zeeman splitting of the ground state is $\abs{\Delta\sub{Z}}/2\pi=\SI{13}{\MHz}$ and the cavity is tuned from $\Delta\sub{C}/2\pi=\SI{-20}{\MHz}$ to $\SI{102.22}{\MHz}$.  Solid lines illustrate the energies of the $\ket{F_x{=}\{0,1\},m_{F_x}{=}0}$ levels, with dashed lines corresponding to $\ketmix{F_x{=}\{0,1\},m_{F_x}{=}0}$.  Below are the full data sets corresponding to three of the data points, showing the driving laser scanned across the cavity resonances found within $\Delta\sub{L}=\Delta\sub{C}\pm 2\Delta\sub{Z}$.  The emission peaks in these traces, (\romnum{1}) to (\romnum{4}), are illustrated and discussed in the text.  Here the detunings of the laser and the cavity are defined such that they are resonant with the $\ket{F_g{=}1,m_{F_g}{=}0} \leftrightarrow \ket{F_x{=}0,m_{F_x}{=}0}$ transition at $\Delta\sub{L}=0$ and $\Delta\sub{C}=0$ respectively.}
\label{fig:ScatteringRates}
\end{figure} 

Direct measurement of the modified coupling strengths predicted by the hyperfine state-mixing is challenging when using the vacuum-stimulated Raman adiabatic passage (V-STIRAP) process for single-photon production as the two transitions that comprise our $\Lambda$-system are oppositely strengthened and weakened by the external field (see \cref{fig:Theory}(c)).  Instead we can infer the relative strengths of these transitions by measuring the polarisation of photons scattered from the driving laser into the cavity by the atom as shown in \cref{fig:ScatteringRates}.  Driving the system with the pump laser at different detunings from the cavity resonance results in four emission peaks.  When the laser is resonant with the cavity the cycling peaks, (\romnum{2}) and (\romnum{3}), are high as they leave the atom in the same magnetic sublevel after the emission of a photon, allowing for many consecutive emissions.  In contrast the Raman resonance peaks, (\romnum{1}) and (\romnum{4}), transfer the atom between different magnetic sublevels, which, for a given detuning of the driving laser, prevents a single atom from emitting multiple photons.  It is obvious that sequential application of the processes (\romnum{1}) and (\romnum{4}) should produce a stream of single-photons alternating between $\sigma^{+}$ and $\sigma^{-}$ polarisation.

Atoms entering the cavity are assumed to be equally distributed across the magnetic substates of the ground level and so the relative heights of these cycling peaks allows us to extract the relative efficiencies of each process.  The upper plot in \cref{fig:ScatteringRates} shows the relative strengths of the cycling transitions for each polarisation as the cavity offset is moved from below the $\ketmix{F_x{=}0}$ level to above $\ketmix{F_x{=}1}$.  When the cavity frequency is close to the $\ket{F_g{=}1} \leftrightarrow \ketmix{F_x{=}0}$ transition the $\sigma^{-}$ emission corresponding to the cycling process between $\ket{F_g{=}1,m_{F_g}{=}1} {\leftrightarrow} \ketmix{F_x{=}0,m_{F_x}{=}0}$ is prevalent.  Conversely when operating near the $\ketmix{F_x{=}1}$ line, $\sigma^{+}$ emission from the $\ket{F_g{=}1,m_{F_g}{=}{-}1} {\leftrightarrow} \ketmix{F_x{=}1,m_{F_x}{=}0}$ transitions dominates.  This behaviour agrees with the nonlinear modification in transition strengths shown in \cref{fig:Theory}(c).

For comparison, if oppositely polarised transitions remained equally coupled then a cavity tuned to resonance with the excited level would result in equally detuned cycling processes and so no preferential emission of either $\sigma^{+}$ or $\sigma^{-}$ photons would occur, disregarding only the effects of weak coupling to other hyperfine levels.  The clear departure from this behaviour is compelling evidence that state-mixing in the intermediate field regime does indeed result in asymmetric coupling strengths between magnetic sublevels.

\subsection{V-STIRAP on the D$_2$ line}

\subsubsection{Single-photon production}

Proper preparation of the initial state is mandatory for any controlled single-photon generation.  For doing so, we only consider those emission processes which immediately follow a single-photon detection in the preceding driving interval.  When a $\sigma^{\pm}$ photon is observed, the atom is considered to have been well prepared in the $\ket{F{=}1,m_F{=}{\mp} 1}$ state and, as such, the next driving interval, when production of the opposite polarisation of photon is attempted, is a good test of emission efficiency.  Considering these correlated emissions we can define $\pcond{\sigma^{\pm}}{\sigma^{\mp}}$ as the probability with which a $\sigma^{\pm}$ photon is detected following the detection of a $\sigma^{\mp}$ photon in the previous driving interval and, it follows, as the efficiency of the $\sigma^{\pm}$ driving process.

\begin{figure}
\includegraphics{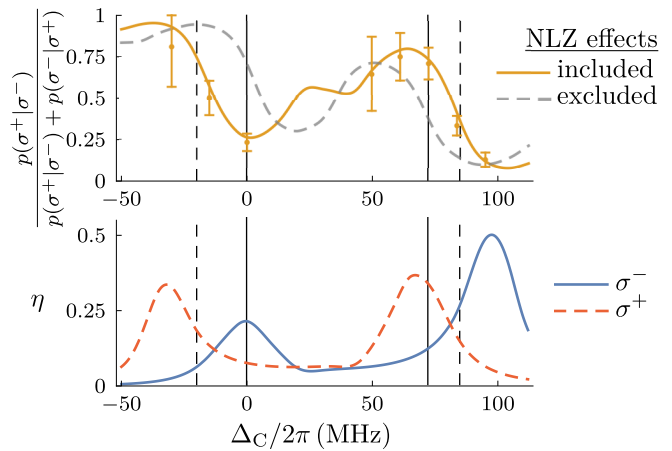}
\caption{The lower plot shows the predicted efficiencies of polarised photon production for each driving process with the system parameters as described in the text.  From these we can infer the expected imbalance in the detection probability of each polarisation, conditioned on the detection of a single photon of the opposite polarisation in the previous driving interval, as described in the text.  This is shown in the upper plot, along with the measured values.  The dashed trace in the upper plot shows the expected behaviour if the NLZ effects are neglected.  The solid vertical lines in both plots illustrate the energies of the $\ket{F_x{=}\{0,1\},m_{F_x}{=}0}$ levels, with dashed lines corresponding to $\ketmix{F_x{=}\{0,1\},m_{F_x}{=}0}$.}
\label{fig:DirectionalEfficiencies}
\end{figure}

 The relative efficiencies which each polarisation of photon is produced are shown in \cref{fig:DirectionalEfficiencies} as the system is operated across the span of the $\ketmix{F_x{=}0}$ to $\ketmix{F_x{=}1}$ levels.  For these experiments a ground state Zeeman splitting of $\abs{\Delta\sub{Z}}/2\pi=\SI{14}{\MHz}$ is applied.  The driving pulse is \SI{500}{\ns} long with a $\sin^{2}$ amplitude profile peaking at $\overline{\Omega_0}/2\pi=\SI{14}{\MHz}$.  The observed asymmetry at each point is in good agreement to the theoretical predictions when the NLZ effects are included in the model.  When the cavity is near resonance with an excited level, the process with an enhanced cavity coupling is more efficient and, as has already been demonstrated, each level strengthens opposite polarisations.  For reference the expected behaviour neglecting the NLZ effects is also included.
 
To reinforce the importance of including these effects, we revisit the experiment of Rempe \etal{} \cite{wilk07,wilk08}: at $\Delta\sub{C}/2\pi{=}\SI{72}{\MHz}$ an imbalance of $\pcond{\sigma^{+}}{\sigma^{-}} / \left\{ \pcond{\sigma^{+}}{\sigma^{-}} + \pcond{\sigma^{-}}{\sigma^{+}} \right\} {\approx} 0.76$ was found, whilst an expected value of ${\approx} 0.41$ was quoted from their simulations.  These two numbers correspond very well with the predicted behaviour of our model including and neglecting the NLZ effects arising from the mixing of magnetic sublevels, respectively.

\begin{figure*}
\centering
\includegraphics{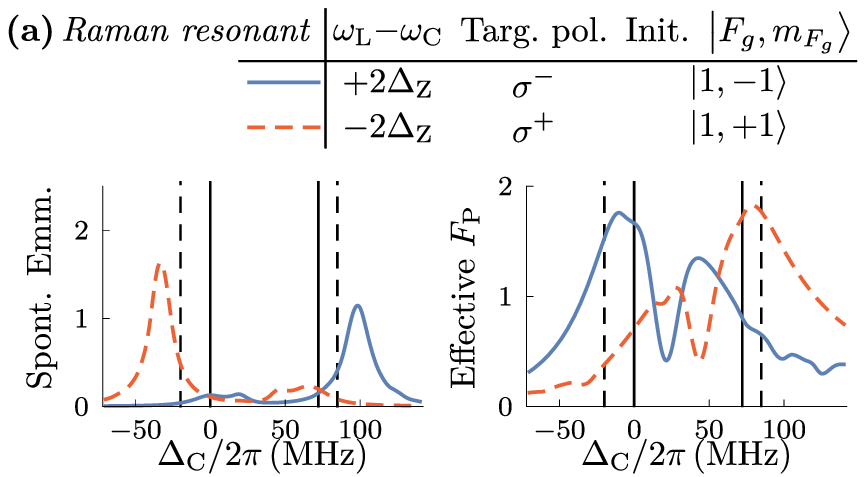}
\includegraphics{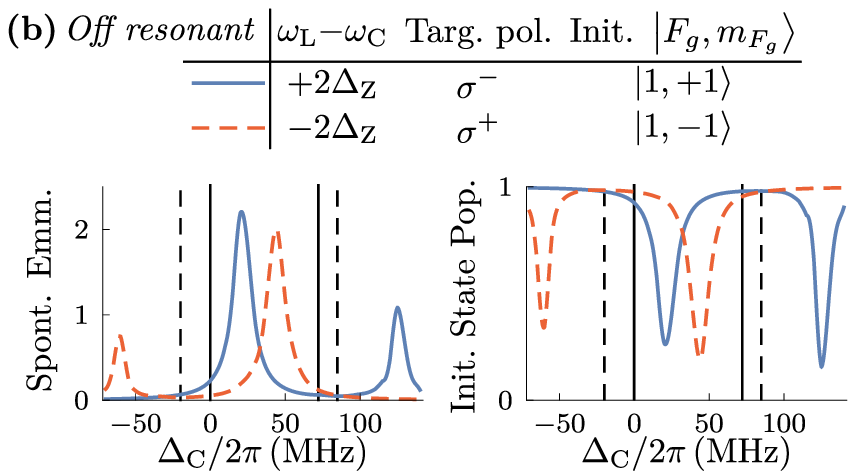}
\caption{(a) The predicted spontaneous emission, in terms of photon number, and effective Purcell factor for the driving processes corresponding to both polarisations of single photon production, as a function of the cavity detuning.  (b) The predicted depopulation of the initial state and spontaneously emitted photon number when the system is driven using the non-Raman resonant laser light that corresponds to the single photon production scheme from the opposite stretched state to that in which the atom is prepared.  The solid vertical lines in both plots illustrate the energies of the $\ket{F_x{=}\{0,1\},m_{F_x}{=}0}$ levels, with dashed lines corresponding to $\ketmix{F_x{=}\{0,1\},m_{F_x}{=}0}$.}
\label{fig:D2DrivingLimitations}
\end{figure*}

It can also be seen that the overall efficiency of single-photon production is higher near the  $\ketmix{F_x=1}$ level than when near $\ketmix{F_x=0}$.  This can be ascribed to the reduced asymmetry between the orthogonal coupling strengths of our $\Lambda$-system in this regime.  To see this experimentally we can compare the relative prevalence of the side-lobes corresponding to our V-STIRAP processes in \cref{fig:ScatteringRates}.  This explains why the system has been previously observed to work better in the $\ketmix{F_x=1}$ region \cite{wilk07,wilk08} and indicates that finding regimes where the coupling strengths are as symmetric as possible is important to efficiently operating the atom-photon interface.

Equal emission of each polarisation is predicted at $\Delta\sub{C}/2\pi=$\SIlist{-13;19.5;81}{\MHz}, with the latter having the highest overall efficiency and so appearing to be the optimal point at which to drive this system.  However production efficiency is only one desirable metric, of equal importance is the suppression of spontaneous emission to ensure high efficiency mutual coherence and indistinguishability of the photons.  This requires high cooperativities, defined as~\cite{kuhn10}
\begin{equation}
	C = \frac{g^2}{2\kappa\gamma}.
	\label{eq:cavCooperativity}
\end{equation}
For an excited atom coupled to a cavity, the ratio of photon emission into the cavity mode versus spontaneous emission into free space is given by the Purcell factor, defined as $2C$ \cite{fox06}.  Here, as we drive a Raman process which in the adiabatic limit never excites the atom out of the ground state, we instead define the effective Purcell factor, $F\sub{P}$, as the ratio of total emission into the cavity versus emission into free space, independent of the atomic state.  It is important to note that this ratio is dependent on the photon production scheme driven, and so is not equivalent to twice the cooperativity of the cavity.

The spontaneous emission and effective Purcell factors predicted by the simulation for the driving processes under consideration are shown in \cref{fig:D2DrivingLimitations}(a). For both processes, spontaneous emission peaks when the pump laser resonantly couples the initial ground state of the atom to either of the excited levels considered.  However, about each level the relative size of these peaks differ greatly between the two processes.  This is again a result of NLZ effects with the asymmetric coupling strengths of oppositely polarised transitions from $\ket{F_g{=}1,m_{F_g}{=}{\pm}1}$ to each excited level resulting in an asymmetric atomic coupling strengths between each arm of the V-STIRAP $\Lambda$-system.  The two possible processes then correspond to having the cavity on the more strongly or weakly coupled transition (with the laser correspondingly coupling the the oppositely polarised, and thus weakened or strengthened, transition).  When the cavity coupling is weakened, as is the case for producing $\sigma^{+}$ ($\sigma^{-}$) photons primarily using the couplings to the $\ketmix{F_x{=}0,m_{F_x}{=}0}$ ($\ketmix{F_x{=}1,m_{F_x}{=}0}$) level, the rate of spontaneous emission correspondingly increases.  It then follows that, around each excited level, the process where the cavity transition is strengthened has a higher effective Purcell factor, which is also shown in \cref{fig:D2DrivingLimitations}(a).  Producing a stream of alternately polarised photons while operating the system around an atomic resonance is then less favourable because for one of the two polarisations, the atom-cavity coupling is substantially weakened.

However the effective Purcell factors for both driving processes are equal when driving between the two excited levels.  This can be understood as the constructive interference of the transition probabilities from each excited level being increased by the strong cavity couplings.  Essentially one can consider both polarisations to be predominantly emitted by the V-STIRAP process using the excited level that provides a strengthened cavity transition.  In the extreme case one could even consider alternating the cavity detuning such that we operate near to resonance with the excited level that efficiently meditates the desired photon production process in both directions.  However, physically this is impractical as the cavity length can not be changed on the required timescales, moreover changing the cavity resonance would result in different frequencies, and thus inherent distinguishability, between the emitted photons.

However, even operating with the cavity resonance tuned between the excited levels is flawed as can be seen in \cref{fig:D2DrivingLimitations}(b) where we consider the effect of the non-Raman resonant polarisation component of the pump laser.  Ideally when the atom is sitting in the wrong stretched state of the ground level for the driving process, the atom would be unaffected.  However with the cavity detuned from resonance with the excited levels, the off-Raman-resonant terms are very efficient at exciting the atomic population out of these states.  Whilst the desired driving scheme works with equal efficiencies and relatively high effective Purcell factors, the final state of these processes would be quickly depleted by the other polarisation component of the same pump laser that produced the single photon.   This inhibits the efficient preparation of the atom in the required state for the following driving process and so the system is not able to efficiently produce streams of alternately polarised photons from a single atom.

With the cavity frequency tuned directly between the two atomic resonances the initial state population of an atom, sitting in the wrong stretched state of the ground level for the driving process, is depleted by less than \SI{25}{\percent} by the non-Raman resonant polarisation component of the pump pulse.  However, referring back to \cref{fig:DirectionalEfficiencies}, single-photon production from the desired process in this regime is even weaker.  Increasing the driving power to produce photons more efficiently also strengthens the off-Raman-resonant excitations, increasing the rate of depopulation of the desired $m_F$ states.  Although NLZ effects increase the separation between the $\ketmix{F_x{=}0}$ and $\ketmix{F_x{=}1}$ levels as the external fields get stronger, the increased Zeeman splitting of the ground level would require further detuned driving processes to account for this.  It is evident that for a cavity of realistic linewidth, the magnitude of magnetic field required to allow it to selectively couple magnetic substates within the \Rb{} \DTwo{} line will always create asymmetries and, ultimately, restrict the system to rather modest efficiencies.

\subsubsection{Two-photon interference}
\label{sec:twoPhotonInterference}

\begin{figure*}
\includegraphics{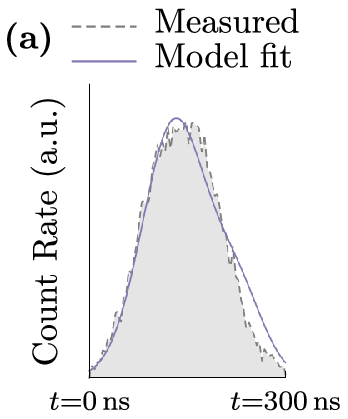}
\includegraphics{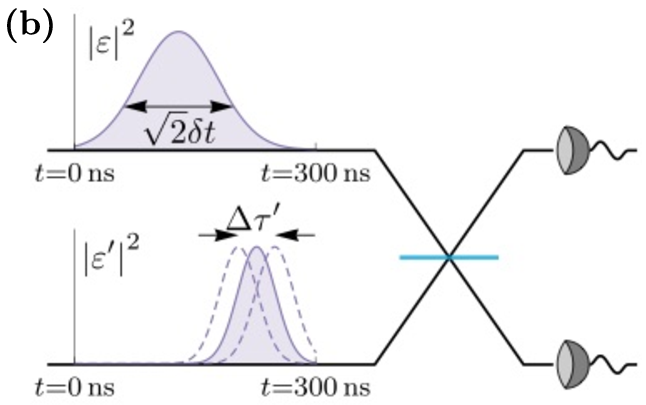}
\includegraphics{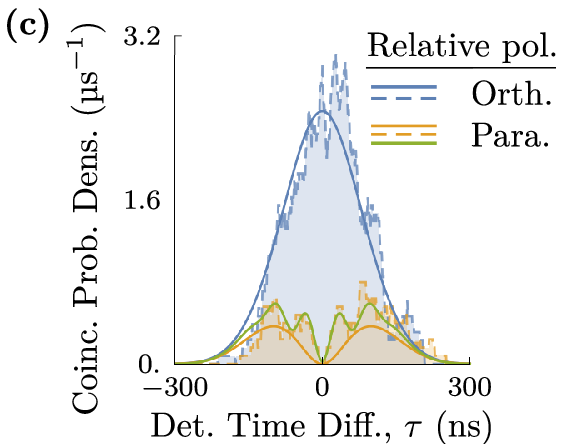}
\caption{The two-photon interference of photons produced using a \SI{300}{\ns} long driving pulse with a $\sin^{2}$ amplitude profile, peaking at $\overline{\Omega\sub{0}}/2\pi=\SI{19}{\MHz}$, a $\abs{\Delta\sub{Z}}/2\pi=\SI{14}{\MHz}$ ground state Zeeman splitting.  The measured intensity profile of the photons, (a), is the combination of photons emitted both as intended and after a spontaneous decay of the atom during the driving process.  The model discussed in \Cref{app:PhotonsContaminatedBySpontaneousEmission}, and fitted as the dashed line, allows typical profiles of photons from these two emission scenarios to be inferred.  These are shown in (b) as the inputs to a HOM experiment, with $\varepsilon$ and $\varepsilon\dash$ corresponding to clean and contaminated emissions respectively.  The measured correlation probabilities of the HOM are shown in (c) (dashed traces) and compared to the behaviour expected when considering only the contaminating effects of spontaneous emission on photon production (orange and blue solid traces) -- the model is described in Appendix B -- for both parallel (indistinguishable) and orthogonally (distinguishable) polarised photon pairs.  The fit to the measured data that includes the contribution of the \SI{9}{\percent} of photon pairs with a frequency beat is also shown (green trace).}
\label{fig:SpontaneousEmissionVsHOM}
\end{figure*}

Spontaneous emission is not only undesirable because it reduces the overall efficiency of the source, but also because it is an incoherent process and thus detrimental to the use of the atom-cavity system as an interface between quantum states of light and matter.  To interrogate this assertion we directly measure the quantum interference of single-photons, produced from the atom-cavity system, using a Hong-Ou-Mandel (HOM) experiment \cite{hong86,legero04}.  Any decoherence of the atom-light system during a photon emission, such as a spontaneous decay resetting the evolution of the atomic state, manifests as a distinguishability between the quantum states of the emitted photons and thus a reduced visibility of the HOM dip.   Using the modelled performance of our source, a back-of-the-envelope calculation approximates the expected impact of spontaneous emissions on the coherence of the emitted photons, which can be compared to the measured HOM visibility.

When pairs of sequentially emitted photons are simultaneously sent into different modes of a 50:50 beam splitter, indistinguishable photons `bunch' and exit in the same output mode, whereas fully distinguishable photons are randomly routed. With the distinguishability of the photon pairs controlled by rotating their relative polarisations, the two-photon visibility is then defined as the reduction in likelihood of both output modes containing a photon when comparing parallel (indistinguishable) to orthogonally (distinguishable) polarised pairs.  The measured probability of recording correlated detections across the beam splitter output ports is shown as a function of detection time difference, $\tau$, in \Cref{fig:SpontaneousEmissionVsHOM}(c) for both these cases.

The \SI{300}{\ns} long photons display an overall two-photon visibility of \SI{70.8\pm4.6}{\percent}, which increases to $\geq\SI{97.8}{\percent}$ when considering only detections within less than $\SI{23}{ns}$ of each other.  The form of this temporal decoherence can be used to infer the coherence properties of the photons \cite{legero06}.  Here, it indicates the interference of narrowband photons with a $2\pi{\times}\SI{2.15}{\MHz}$ bandwidth with \SI{91}{\percent} of the signal from photon pairs of the same frequency, and the remaining contribution from pairs exhibiting a frequency difference of ${\sim}2\pi{\times}\SI{15}{\MHz}{\sim}\abs{\Delta\sub{Z}}$ which arises from undesired driving processes between the magnetic sublevels.  This is shown as a superimposed beat signal (green trace) in \cref{fig:SpontaneousEmissionVsHOM}(c).  Disregarding this secondary process the fundamental visibility of the narrowband photons produced as intended is then \SI{77.4}{\percent}.

To consider how spontaneous emission effects the two-photon visibility we approximate all photons to have a Gaussian intensity envelope and the probability of spontaneous emission to be proportional to the intensity profile of the the driving laser. Typical temporal profiles of photons both affected and unaffected by a spontaneous emission can then be inferred from the measured profile of photon detections.  The recorded photon emission profile is shown in \cref{fig:SpontaneousEmissionVsHOM}(a) along with the best fit of the emission model used, the details of which can be found in \Cref{app:PhotonsContaminatedBySpontaneousEmission}.

The normalised wavepacket of a Gaussian photon can be expressed as
\begin{equation}
	\varepsilon(t,\delta t, t_0) = \left(\frac{2}{\pi \delta t^{2}}\right)^{1/4} \e^{-\left(\frac{t-t_0}{\delta t}\right)^{2}}
	\label{eq:gaussianWavepackets}
\end{equation}
where $\delta t$ is the width parameter and $t_0$ determines the arrival time of the photons peak.  We find photons unaffected by spontaneous emission have $\delta t=\SI{97.4}{\ns}$ and $t_0 = \SI{129.5}{\ns}$.  Those emitted after a spontaneous emission has reset the atom to the ground state have a correspondingly shortened wavepacket and later arrival time.  These two properties can physically be understood to be linked because photon emission can only continue within the \SI{300}{\ns} duration of the pump laser, and thus a later emission necessitates a shorter photon.  The probabilistic nature of these spontaneous decays will result in a variation of these properties, which we account for by considering truncated emissions with a typically width of $\delta t \dash=\SI{47.6}{\ns}$ and with the peak arrival time described by Gaussian distribution centred about $t_0\dash = \SI{226.7}{\ns}$ with a standard deviation of $\Delta \tau \dash / \sqrt{2} = \SI{31.8}{\ns}$.  Both the clean and shortened emission profiles are shown in \cref{fig:SpontaneousEmissionVsHOM}(b) as the input pair to a HOM experiment.


Modelling the two-photon interference of photons of different length, with a jitter in arrival time on top of an overall offset, is an application of the work described in \cite{legero06} by Legero \etal{}  Our approach, which takes this work and then weights each pair-emission scenario -- such as two clean emissions, a clean emission followed by a contaminated emission and so on -- by their relative prevalence in the system, is described in \Cref{app:TwoPhotonInterference}.  The predicted coincidence probability densities   are overlayed, with solid lines, over the measured behaviour, dashed lines, in \cref{fig:SpontaneousEmissionVsHOM}(c).  The two-photon visibility predicted by our model is \SI{80.9}{\percent}.  This is then an approximate upper bound on the visibility these photons could ever be expected to exhibit.

The predictions of the model are consistent with the observed behaviour, both in the limited two-photon visibility and the temporal form this decoherence takes.  The model replicates the dip to near-perfect interference for approaching simultaneous photon detections.  The rate at which the photons decohere can also be seen to be compatible with that measured once we account for the effects of the already discussed frequency difference of \SI{9}{\percent} of photon pairs.  This frequency difference results in the relative phase between the interfering photons running at ${\sim}2\pi{\times}\SI{15}{\MHz}$, changing the HOM interference of otherwise indistinguishable pairs between `bunching' (no cross-detector correlations) to `anti-bunching' (no same-detector correlations) at a commensurate rate.  The result is the observed narrowing of the characteristic HOM dip as $\tau \rightarrow 0$ and an oscillation in the measured correlation rate for increasing $\abs{\tau}$.  This interference behaviour is then overlaid with the interference profile of the remaining \SI{91}{\percent} of photon pairs that were produced from the desired driving scheme.  This remaining temporal decoherence that is not attributable to the frequency beat is where the effects of spontaneous emission are expected to be observed.  Overall, the qualitative agreement between the model and the measurements further endorses the significant impact of spontaneous emission on the coherence of the atom-cavity system.

\subsection{V-STIRAP on the D$_1$ line}

Motivated by the difficulties inherent to operating the polarised atom-cavity system on the \DTwo{} line, we now evaluate the feasibility of using the \DOne{} line, corresponding to the transition $5^2S_{1/2} \leftrightarrow 5^2P_{1/2}$.  This has the advantage of a far greater hyperfine splitting between the two excited levels, $\ket{F_x{=}1}$ and $\ket{F_x{=}2}$, of $2 A_{\mathrm{hfs,}5^2 \mathrm{P}_{1/2}} / \hbar = 2\pi{\times}\SI{816.66}{\MHz}$ \cite{steck87}.  As such the mixing of these hyperfine levels, and so NLZ effects are almost negligible in the regime where we would need to operate the source.  The trade-off is that the couplings are inherently weaker, both from lower $A_{ij}$'s and because the reduced dipole matrix element is weaker at $\redmel{J_g{=}1/2}{e\vec{r}}{J_x{=}1/2} \approx 0.75 {\times} \redmel{J_g{=}1/2}{e\vec{r}}{J_x{=}3/2}$~\cite{steck87}.

\begin{figure*}[t!]
\begin{minipage}[t]{0.25\linewidth}
	\includegraphics{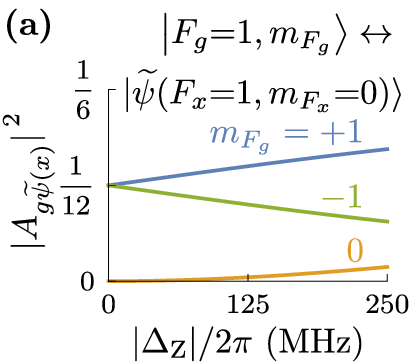}
\end{minipage}%
\begin{minipage}[t]{0.75\linewidth}
\includegraphics{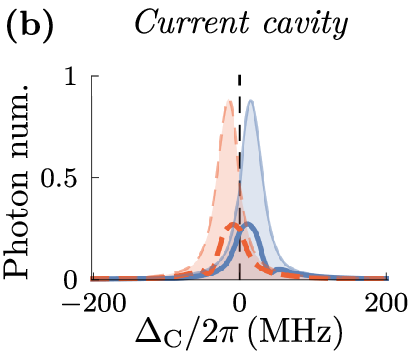}
\includegraphics{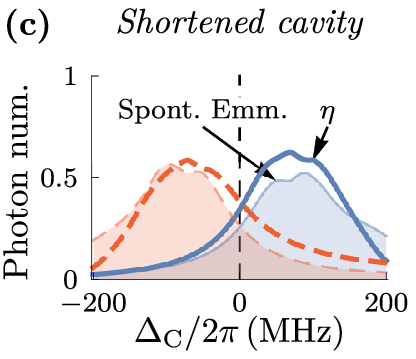}
\includegraphics{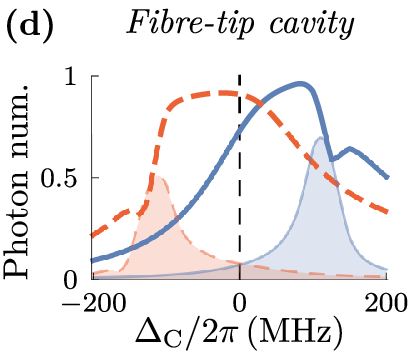}\\[0.15\baselineskip]
\includegraphics{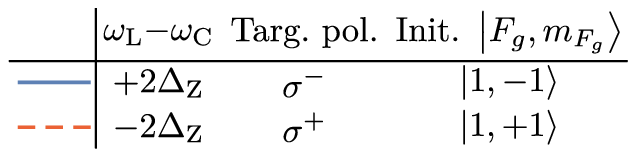}
\end{minipage}

\caption{(a) Polarisation transition strengths for transitions from the magnetic sublevels of the $\ket{F_g={1}}$ state to the $\ket{F_x{=}1, m_{F_x}{=}0}$ excited sublevel expressed as multiples of the \DOne{} transition dipole matrix element squared, $\abs{\redmel{J_g=1/2}{e\vec{r}}{J_x=1/2}}^2$.  The field strength is expressed in terms of the Zeeman splitting it produces in the ground state. (b)-(d) The photon production efficiency (thick, unshaded) and spontaneous emission (thin, shaded) expressed in terms of photon number for driving processes corresponding to both $\sigma^{+}$ and $\sigma^{-}$ production.  The performance in the region of the $\ket{F_x={1},m_{F_x}=0}$ sublevel of the \DOne{} line (vertical dashed line) is shown for the current cavity, (b), an equivalent cavity of $1/5$ the length, (c), and a fibre-tip cavity, (d), the parameters for which are described in the text.}
\label{fig:D1Driving}
\end{figure*}

The modified coupling strengths when utilising the $\ket{F_g{=}1,m_{F_g}} \leftrightarrow \ \ketmix{F_x{=}1,m_{F_x}{=}0}$ transitions are shown in \cref{fig:D1Driving}(a).  The strengths of oppositely polarised transitions are oppositely modified, however, at equivalent field strengths, this effect is considerably smaller in comparison to the \DTwo{} line.  Couplings to the $\ketmix{F_x{=}2,m_{F_x}{=}0}$ sublevel are not shown as the system is best operated around the $\ketmix{F_x{=}1}$ level.  This is because the stretched states, $\ketmix{F_x{=}2,m_{F_x}{=}{\pm}2}$, are coupled to the ground $\ketmix{F_g{=}1,m_{F_{g}}{=}{\pm}1}$ sublevels by the pump laser, resulting in cycling processes and rapid depopulation of the desired states.

To compensate for the inherently weaker couplings of the \DOne{} line, a cavity that provides stronger coupling to the atom is required.  To see this consider \cref{fig:D1Driving}(b), which shows the production of single-photons that would be achievable when using our present system with $\{\overline{g},\kappa,\gamma\}/2\pi = \{7.27,1.875,3\}\,\si{\MHz}$ -- where the atom-cavity coupling rate has been recalculated using the reduced dipole matrix element for the \DOne{} line.  The system once again has an applied ground state Zeeman splitting of $\abs{\Delta\sub{Z}}/2\pi=\SI{14}{\MHz}$, but the pump intensity has been increased to $\overline{\Omega_0}/2\pi=\SI{35}{\MHz}$ as this is needed to stimulate any significant photon emission.  The efficiency with which each polarisation is produced is nearly symmetric about the excited level due to the state mixing being essentially negligible, however spontaneous emission far exceeds photon production.  This is unsurprising as the zero-field atom-cavity coupling from the stretched ground sublevels to $\ketmix{F_x{=}1}$ is $g = \overline{g} / \sqrt{12} =2\pi{\times}\SI{2.12}{\MHz}$.  Using \cref{eq:cavCooperativity} this gives a cooperativity of $C=0.40$, below the $C > 1$ that typically defines the strong-coupling regime. In such conditions it is inevitable that spontaneous emission will be significant in comparison to single-photon production.

The square of the atom-cavity coupling rate is inversely proportional to the volume of the cavity mode, $g \propto V\sub{m}^{-1/2}$~\cite{kuhn10}, and so reducing the mode volume is key to increasing the interaction strength.  A physically realistic example is to consider the case where our cavity is shorted to $\SI{339}{\um}/5 = \SI{67.8}{\um}$.  The system parameters would then become $\{\overline{g},\kappa,\gamma\}/2\pi = \{24.29,9.375,3\}\si{\MHz}$ with a zero-field atom-cavity coupling of $g / 2\pi = \SI{7.07}{\MHz}$ and cooperativity of $C \approx 0.89$.   To accommodate for the broader cavity linewidth, $\Delta\omega\sub{FWHM}= 2\kappa$, the Zeeman splitting of the ground state is increased to $\abs{\Delta\sub{Z}}/2\pi=\SI{70}{\MHz}$ and the driving power used to $\overline{\Omega_0}/2\pi=\SI{120}{\MHz}$.  The performance of this system is shown in \cref{fig:D1Driving}(c).  At $\Delta\sub{C}/2\pi=\SI{3.9}{\MHz}$, when the cavity is on resonance with the $\ketmix{F_x{=}1,m_{F_x}{=}0}$ sublevel, $\sigma^{+}$ and  $\sigma^{-}$ photons are produced with effective Purcell factors of \numlist{1.34;1.47} respectively.  Whilst this shortened cavity and the modified driving parameters are not designed to be the optimum point at which to operate on the \DOne{} line, their performance exceeds that which we calculated on the \DTwo{} line, and so demonstrates the improvement that is readily available using by current technologies in an informed manner.  We emphasise that this conclusion is surprising, because previous models which neglect NLZ effects predict that the inherently stronger coupling of the \DTwo{} line allows for more efficient photon production in comparison to the \DOne{} line.

Further improvement could be found in the recent development of fibre-based cavities.  In 2010 Jakob Reichel \etal{} demonstrated a Fabry--P\'{e}rot cavity of finesse $\Fin>\num{130000}$, formed between mirrors ablated to the tips of optical fibres \cite{hunger10}.  The mirror curvatures, and so mode volumes, achievable using this process are significantly beyond the technical limitations of the super-polished glass substrates from which our and previous systems \cite{nisbet11,kuhn02,wilk07} have been constructed.  Since this initial demonstration fabrication methods have further improved \cite{takahashi14} and fibre-tips are now a viable alternative \cite{gallego18}.

Consider a $\SI{128}{\um}$ long fibre-based cavity that has mirror curvatures of $R\sub{cav}=\SI{200}{\um}$, with \SI{10}{ppm} scattering losses from each mirror and transmissions of \SIlist{5;40}{ppm}.  These specifications are typical for fibre-tip mirrors already demonstrated.  This cavity would increase our atom-cavity coupling rates to $\overline{g}/2\pi=\text{\SIlist{66;95}{\MHz}}$ for the \DOne{} and \DTwo{} lines, respectively.  The relatively large linewidth of the cavity, $\omega\sub{FWHM}/2\pi=\SI{12.12}{\MHz}$, when compared to the hyperfine splitting of the energy levels on the \DTwo{} line would once again necessitate a sufficiently strong external magnetic field as to be problematic and so we instead consider driving the process on the \DOne{} line.  For this simulation, shown in \cref{fig:D1Driving}(d), the driving power was $\overline{\Omega_0}/2\pi=\overline{\g}/2\pi=\SI{66}{\MHz}$ and the ground state Zeeman splitting was set to $\abs{\Delta\sub{Z}}/2\pi=\SI{42}{\MHz}$.  Even without any attempt of finding the optimal driving parameters, it is clear that the efficiency and purity of the photon emissions from the new system is predicted to far exceed anything that could be achieved with conventional mirror substrates.  Quantifying this at two points of interest we see that:
		\begin{itemize}
			\item at $\Delta\sub{C}/2\pi=\SI{29}{\MHz}$ both processes have \SI{86.3}{\percent} efficiency and effective $F\sub{P}$'s of 14.1 ($\sigma^{+}$ generation) and 7.86 ($\sigma^{-}$),
			\item at $\Delta\sub{C}/2\pi=\SI{10}{\MHz}$ both processes have effective $F\sub{P}$'s  of 10.85 and are \SI{90.0}{\percent} ($\sigma^{+}$) and \SI{78.3}{\percent} ($\sigma^{-}$) efficient. 
		\end{itemize}
At the same time, spontaneous emission is now negligible, which should lead to photons of excellent coherence properties.
\section{Outlook}

In this paper we have demonstrated the significance of considering NLZ effects to the understanding and operation of a cavity-enhanced polarised atom-light interface. This has been investigated both theoretically and in an experimental setting with a single \Rb{} atom coupled to a high finesse optical cavity.  The modified atomic properties are calculated by numerically diagonalising the Hamiltonian of an atom in a static magnetic field and incorporated into the full atom-cavity system, the evolution of which is then modelled using a master equation approach.  This two-step approach successfully predicts the observed experimental behaviour.

The predicted modification in transition strengths between the ground and excited levels of the \DTwo{} line were directly observed in the relative efficiency with which resonant photons were scattered into the cavity mode.  One direction of the atom-light interface, single-photon production using V-STIRAP, was then considered with the asymmetrically modified couplings of the $\Lambda$-system.  This lead to strongly imbalanced production efficiencies for oppositely polarised photons.


Because atom-photon interfaces are of particular interest to quantum networking applications we measured the coherence properties of sequentially emitted single-photons.  The imperfect interference of these pairs was consistent with a simple approximation of the impact of spontaneous emission within the system.  The increase in these incoherent processes is a predicted result of the modified atomic transitions and strongly suggests that a reliable interface must be operated in a regime that minimises NLZ effects.  Despite the present limitations, polarised photons from our system have recently been used to successfully demonstrate multimode interferometry with cavity-photons \cite{barrett18a}.

The suppressed state-mixing of the \DOne{} line was shown theoretically to provide significant improvements, provided the weaker atomic coupling is counter-balanced by a cavity that provides a stronger coupling.  This demonstrates how the insights gained from considering NLZ effects inform the future design of such systems.  With the ongoing research into tools for realising distributed quantum networks and the desire for a reliable interface between nodal qubits and their photonic links, we anticipate that the results described and measured in this work will pave the way to implementing a highly reliable atom-photon interface in a scalable quantum network. 
\section{Acknowledgements}

The authors would like to acknowledge support for this work through the quantum technologies programme (NQIT hub), and express their gratitude to G.~Rempe for his helpful discussions.

\appendix
\crefalias{section}{appendix}
\section{Photons contaminated by spontaneous emission}
\label[appendix]{app:PhotonsContaminatedBySpontaneousEmission}

\begin{figure}[t]
\includegraphics{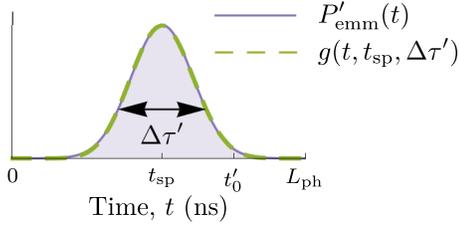}
\caption{The probability, $P\sub{emm}\dash$, of a photon emission into the cavity beginning at time $t$ after a spontaneous emission has reset the atom to the ground state.  This is well approximated by a Gaussian distribution, $g(t,t\sub{sp},\Delta\tau\dash)$.}
\label{fig:ContaminatedPhotonEmmissionTimes}
\end{figure}

To consider the effect of spontaneous emission on photon production we assume that the probability of a spontaneous emission is proportional to the intensity profile of the the driving laser, \ie{} $P\sub{sp}(t) \propto \sin^4(\pi t / L\sub{ph})$, where $L\sub{ph}=\SI{300}{\ns}$.  From this, the probability that at time $t$ a spontaneous emission has occurred at $t\dash<t$ and that no further spontaneous emission occurs for $t\dash>t$ is
\begin{equation}
	P\submath{sp}{<t} (t) \propto \int^{t}_{0} P\sub{sp}(t\dash) \diffd t\dash \times \Big(1 -  \int^{L\sub{ph}}_{t} P\sub{sp}(t\dash) \diffd t\dash \Big).
	\label{eq:probSpontBeforeTOnly}
\end{equation}
The probability that the final spontaneous emission is at time $t$ is then
\begin{equation}
	P\submath{sp}{=t} (t) \propto \frac{\diffd}{\diffd t} P\submath{sp}{<t} (t).
	\label{eq:probSpontFinal}
\end{equation}
Denoting the measured emission profile of the photons as $\abs{\psi}^{2}(t)$ -- this corresponds to the solid trace in \cref{fig:SpontaneousEmissionVsHOM}(a) -- the probability of a final spontaneous emission at time $t$ followed by the emission of a photon from the cavity at some time $t\dash>t$ is
\begin{equation}
	P\dash\sub{emm}(t) \propto P\submath{sp}{=t} (t) \times \int_{t}		^{L\sub{ph}} \abs{\psi}^{2}(t\dash) \diffd t\dash.
	\label{eq:PemmContaminated}
\end{equation}
This can be well approximated by a Gaussian distribution of the form
\begin{equation}
	g(t,t_0,\Delta\tau) = \frac{1}{\Delta\tau \sqrt{\pi}} \exp(-\frac{(t-t_0)^{2}}{\Delta\tau^{2}}),
	\label{eq:Gaussian}
\end{equation}
as is shown in \cref{fig:ContaminatedPhotonEmmissionTimes} with $t_0=t\sub{sp}=\SI{153.4}{\ns}$ and $\Delta\tau=\Delta\tau\dash=\SI{45.0}{\ns}$.

The next step is to take this distribution in the start time of the emission of contaminated photons and approximate their wavepackets.  With the wavepackets of photons unaffected by spontaneous emission defined as in \cref{eq:gaussianWavepackets}, $\varepsilon(t,\delta t,t_0)$, we postulate the affected photon profiles to be
\begin{equation}
	\varepsilon(t,\delta t\dash \equiv \frac{L\sub{ph}-t\sub{sp}}{L\sub{ph}}\delta t ,t_0\dash \equiv \frac{L\sub{ph} + t\sub{sp}}{2})
\end{equation}
Physically this is taking the typical length of the contaminated emission wavepackets to be shortened by an amount corresponding to their later emission, and assuming their profile remains a symmetric Gaussian.  We emphasise that in reality the emission time and wavepacket length of the photons will be codependent and varied but for our purposes we instead approximate them with an average value.  The overall emission profile we expect to measure is then the weighted sum of the two emission cases
\begin{equation}
\begin{aligned}
	\abs{\psi\sub{model}}^{2} (t) = (1-P\sub{cont})\abs{\varepsilon(t,\delta t, t_0)}^{2} + &\\
	P\sub{cont} \int_{-\infty}^{\infty} g(t, \Delta t, \Delta \tau) \abs{\varepsilon(t,\delta t\dash, t_0\dash + \Delta t)}^{2} \diffd \Delta t &,
\label{eq:modelEmissionProfile}
\end{aligned}
\end{equation}
where $P\sub{cont}$ is the probability that an cavity emission is contaminated by a prior spontaneous emission and we have averaged over the jittered emission time of the contaminated events.

To find $P\sub{cont}$ we first recall that the ratio of emission into the cavity and spontaneous emission defines an effective Purcell factor, $F\sub{P}$, and thus take the proportion of total emission that is uncontaminated by spontaneous emission to be $F\sub{P} / (F\sub{P}+1)$.  For a photon emission to follow a spontaneous emission we insist that the atom must decay into the correct Zeeman sublevel of $\ket{F_g{=}1}$.  Considering the shifted coupling strengths of the various decay channels within the atom we can define the probability that the decay from the excited level will be to the ground $\ket{F_g,m_{F_g}}$ sublevel as $P\submath{decay,}{\ket*{F_g,m_{F_g}}}$.  Finally we weight the relative contribution of their produced photons of each polarisation by the overall production efficiency, $\eta$.  Overall this gives
\begin{equation}
\begin{aligned}
	P\sub{cont} = 
	&\frac{\eta_{\sigma^{+}}}{\eta_{\sigma^{+}} {+} \eta_{\sigma^{-}}} P\submath{decay,}{\ket*{1,+1}} \left( 1 - \frac{F\submath{P}{\sigma^{+}}}{F\submath{P}{\sigma^{+}}{+}1}\right) \\
	+ &\frac{\eta_{\sigma^{-}}}{\eta_{\sigma^{+}} {+} \eta_{\sigma^{-}}} P\submath{decay,}{\ket*{1,-1}} \left( 1 - \frac{F\submath{P}{\sigma^{-}}}{F\submath{P}{\sigma^{-}}{+}1} \right).
	\label{eq:Pcont}
\end{aligned}
\end{equation}

The model in \cref{eq:modelEmissionProfile} can then be fitted to the measured photon emission profile.  Overall applying this procedure to our system gives the parameters of the two photon emission profiles as $\{ \delta t, t_0 \} = \{ 97.4, 47.6 \}\,\si{\ns}$ and $\{\delta t\dash, t_0\dash, \Delta\tau\dash \} = \{ 47.6, 226.7, 45.0 \}\,\si{\ns}$.  , The fitted model and emission profiles are shown in \cref{fig:SpontaneousEmissionVsHOM}(a) and (b) respectively.

\section{Two-photon interference}
\label[appendix]{app:TwoPhotonInterference}

For two photons input into a HOM experiment the probability of recording a correlated detection, separated in time by $\tau$, between the output modes varies with the distinguishability of the input states.  In the limiting cases of indistinguishable, $P_\para$, and orthogonal, $P_\perp$ input pairs this is given by \cite{legero03,legero06}
\begin{align}
	P_\perp(\tau) &= \frac{1}{2} (\abs{\varepsilon_A}^{2} * \abs{\varepsilon_B}^{2})(\tau), \\
	P_\para(\tau) &= P_\perp(\tau) - \int_{\infty}^{\infty} F(t,\tau) \diffd t,
\end{align}
where
\begin{equation}
	F(t,\tau) = \frac{1}{2}\varepsilon_A(t)\varepsilon_A(t+\tau)\varepsilon_B(t)\varepsilon_B(t+\tau).
\end{equation}
Here we have implicitly made the assumption that the photons remain in phase, thus the correlation probabilities are functions of only the photon wavepacket amplitudes.  For the Gaussian wavepackets we consider we define
\begin{align}
	\varepsilon_A(t)&=\varepsilon(t,\delta t_A,\frac{\delta\tau\sub{off}+\delta\tau}{2}), \\
	\varepsilon_B(t)&=\varepsilon(t,\delta t_B,-\frac{\delta\tau\sub{off}+\delta\tau}{2}), 
\end{align}
where $\delta t_A$ and $\delta t_B$ correspond to each photons length and the offset in arrival times has both a fixed, $\delta\tau\sub{off}$, and varying, $\delta\tau$, component.  Taking the jitter in the arrival time difference to follow a Gaussian, $g(t,t-\delta \tau,\Delta \tau)$ in the form of \cref{eq:Gaussian}, the modified correlation probabilities are
\begin{equation}
	P_{\perp,\para}(\tau) = \int_{\infty}^{\infty} g(t,t-\delta \tau,\Delta \tau) P_{\perp,\para}(\tau) \diffd \delta\tau.
\end{equation}
For our Gaussian photons this gives analytical solutions, however these do not simplify to an readily presentable form.

The four possible pair emission scenarios -- corresponding to the clean or contaminated emission of $\sigma^{+}$ and $\sigma_{-}$ photons -- are then weighted by their relative prevalence in the system.  This is done using analogous arguments to those used to define the overall probability that a single emission is contaminated, \cref{eq:Pcont}.  The final prediction for the interference we expect to see is shown as the solid traces in \cref{fig:SpontaneousEmissionVsHOM}(c).


\bibliographystyle{apsrev4-1}
\bibliography{NLZ_References}

\end{document}